\documentclass[5p, times, twocolumn]{elsarticle}
\usepackage{subfigure}

\usepackage{graphicx}
\usepackage{amssymb, amsbsy,latexsym}
\usepackage{lineno}

\journal{Nuclear Instruments and Methods A}

\begin{document}

\begin{frontmatter}



\title{Adaptation of frequency-domain readout for 
Transition Edge Sensor bolometers 
for the POLARBEAR-2 Cosmic Microwave Background experiment}

% if there is only one institution, use this form:
%\author{John Author, Giovanna Autore}
%\address{University of Wisdom, Physics City, Scienceland}

% else, use optional labels to link authors explicitly to addresses,
% as shown below:
\author[A]{Kaori Hattori}
\author[C]{Kam Arnold}
\author[C]{DarcyBarron}
\author[D]{Matt Dobbs}
\author[D]{Tijmen de Haan}
\author[B]{Nicholas Harrington}
\author[A]{Masaya Hasegawa}
\author[A]{Masashi Hazumi}
\author[B]{William L. Holzapfel}
\author[C]{Brian Keating}
\author[B]{Adrian T. Lee}
\author[A]{Hideki Morii}
\author[B]{Michael J. Myers}
\author[E]{Graeme Smecher}
\author[B]{Aritoki Suzuki}
\author[A]{Takayuki Tomaru}
%\author[C]{Peter Ade}
%\author[L]{Julian Borrill}
%\author[E]{Scott Chapman}
%\author[A]{Yuji Chinone}
%\author[F]{Josquin Errard}
%\author[G]{Giullo Fabbian}
%\author[B]{Adnan Ghribi}
%\author[C]{William Grainger}
%\author[H]{Nils Halverson}
%\author[I]{Yuki Inoue}
%\author[J]{Sou Ishii}
%\author[K]{Yuta Kaneko}
%\author[D]{Brian Keating}
%\author[B]{Zigmund Kermish}
%\author[A]{Nobuhiro Kimura}
%\author[L]{Ted Kisner}
%\author[B]{William Kranz}
%\author[D]{Frederick Matsuda}
%\author[A]{Tomotake Matsumura}
%\author[B]{Haruki Nishino}
%\author[A]{Takahiro Okamura}
%\author[B]{Erin Quealy}
%\author[B]{Christian L. Reichardt}
%\author[B]{Paul L. Richards}
%\author[B]{Darin Rosen}
%\author[E]{Colin Ross}
%\author[I]{Akie Shimizu}
%\author[B]{Mike Sholl}
%\author[D]{Praween Siritanasak}
%\author[E]{Peter Smith}
%\author[D]{Nathan Stebor}
%\author[G]{Radek Stompor}
%\author[A]{Jun-ichi Suzuki}
%\author[J]{Suguru Takada}
%\author[A]{Ken-ichi Tanaka}
%\author[L]{Oliver Zahn}
\address[A]{High Energy Accelerator Research Organization (KEK), Tsukuba, Japan}
\address[B]{University of California, Berkeley, Physics, Berkeley, CA, USA}
%\address[C]{Cardiff University, The Parade, Cardiff, UK}
\address[C]{University of California, San Diego, La Jolla, CA, USA}
%\address[E]{Dalhousie University, Halifax, Canada}
\address[D]{McGill University, Montreal, Quebec, Canada}
%\address[G]{Laboratoire Astroparticule et Cosmologie (APC), Universite Paris 7, Paris, France}
%\address[H]{University of Colorado, Boulder, CO, USA}
%\address[I]{The Graduate University for Advanced Studies, Tsukuba, Japan}
%\address[J]{University of Tsukuba, Tsukuba, Japan}
%\address[K]{University of Tokyo, Tokyo, Japan}
%\address[L]{Lawrence Berkeley National Laboratory, Berkeley, CA, USA}
\address[E]{Three-Speed Logic, Inc., Vancouver, Canada}

\begin{abstract}
The POLARBEAR-2 Cosmic Microwave Background (CMB) experiment aims
to observe B-mode polarization with high sensitivity
to explore gravitational lensing of CMB and inflationary gravitational waves.
POLARBEAR-2 is an upgraded experiment based on POLARBEAR-1, which had
first light in January 2012.
For POLARBEAR-2, we will build a receiver that has 7,588 
Transition Edge Sensor (TES) bolometers coupled
to two-band (95 and 150 GHz) polarization-sensitive antennas.
For the large array's readout, we employ digital frequency-domain 
multiplexing and multiplex 32 bolometers 
through a single superconducting quantum interference device (SQUID).
An 8-bolometer frequency-domain multiplexing readout has been deployed 
on POLARBEAR-1 experiment. Extending that architecture to 32 bolometers 
requires an increase in the bandwidth of the SQUID electronics to 3 MHz. 
To achieve this increase in bandwidth, we use Digital Active Nulling (DAN)
on the digital frequency multiplexing platform.
In this paper, we present requirements and improvements on
parasitic inductance and resistance of cryogenic wiring and capacitors
used for modulating bolometers.
These components are problematic above 1 MHz.
We also show that our system is able to bias a bolometer 
in its superconducting transition at 3 MHz.

\end{abstract}

\begin{keyword}
TES bolometer
\sep
digital feedback
\sep
frequency-domain multiplexing
\sep
Cosmic Microwave Background
\sep
POLARBEAR-2

%% Find PACS codes here: http://www.aip.org/pacs/pacs2010/individuals/pacs2010_regular_edition/index.html


\end{keyword}

\end{frontmatter}

%\linenumbers

\section{Introduction}

The POLARBEAR experiment \cite{ref:polarbear} aims 
to observe the B-mode polarization pattern
imprinted on the cosmic microwave background (CMB).
The B-mode polarization, not discovered yet, 
is a powerful tool for 
exploring gravitational lensing of CMB and inflationary gravitational waves
\cite{ref:B-mode}.
The POLARBEAR-1 experiment had first light in
January 2012 at 5150 meters altitude in the Atacama Desert
\cite{ref:PB1} \cite{ref:PB1-2}.
Scientific observations have been ongoing with 1,274 
single band (150 GHz) antenna-coupled, 
polarization sensitive Transition Edge Sensor (TES) bolometers.

POLARBEAR-2 is an upgraded experiment, which will be more sensitive
to the B-mode and E-mode polarization signals \cite{ref:PB2}.
We will build a receiver that has 7,588 TES bolometers coupled
to two-band (95 and 150 GHz) polarization-sensitive antennas \cite{ref:PB2-antenna}. 
The kilopixel arrays of multi-band polarization-sensitive pixels
are necessary to achieve the high sensitivity required by these science goals. 
Multiband observations were chosen to mitigate the foreground
produced by synchrotron radiation and dust emission.

For TES bolometer array readout, 
we employ digital frequency-domain multiplexing 
(fMUX) \cite{ref:dfmux} and multiplex bolometers through 
a single superconducting quantum interference device (SQUID)
mounted on the 4 K stage.
Each TES bolometer ($R_{bolo}$) is connected to an inductor ($L$)
and a capacitor ($C$) in series to define a frequency 
of its voltage-bias carrier,
shown in Fig. \ref{fig:diagram_pb1}.
An 8-bolometer fMUX readout has been deployed 
on the POLARBEAR-1 experiment.
The fMUX readout has been also deployed on 
the APEX-Sunyaev-Zel’dovich (SZ) \cite{ref:APEX-SZ} and 
South Pole Telescope (SPT) SZ \cite{ref:SPT-SZ} which have 7x multiplexing,
SPT pol with 12x \cite{ref:SPT-pol}, and EBEX with 16x \cite{ref:EBEX}.
Polarbear-2 will employ 32x multiplexing 
to read out its large number of detectors.
In the following section, we 
describe the requirements imposed by this high multiplexing factor.

\section{Improvements for readout}

\subsection{Bandwidth}

Extending that architecture to 32 bolometers requires 
an increase in the bandwidth of the SQUID electronics. 
The bandwidth is determined by spacing of resonances
defined by LC filters.
%The spacing is determined by timeconstant of bolometers.
The spacing is chosen such that cross talk
from adjacent resonance is small in comparison
to the other sources of cross talk in the experiment.
For POLARBEAR-2 a spacing of 90 kHz will be used
and the total bandwidth will be 3 MHz.
The spacing is similar to that chosen for POLARBEAR-1.
%The bandwidth of the readout used for POLARBEAR-1 is limited to about 1 MHz, 
%due to phase delay from the 300 K to 4 K wiring 
%used for the SQUID flux locked loop.
Because of the limited dynamic range of the SQUID, 
the POLARBEAR-1 readout used a broadband analog feedback loop \cite{ref:afmux}, 
which is limited to 1 MHz due to phase delay from the 300 K to 4 K wiring.
To overcome this limitation, we use an alternative
feedback technique called Digital Active Nulling (DAN) \cite{ref:DAN} 
on the digital frequency multiplexing platform.
With DAN, digital feedback is
calculated across the narrow bandwidth of each bolometer, 
extending the useful bandwidth of the SQUID amplifier.

%\subsection{Frequency spacing}
%The inductance of $L=$24 $\mu$H was chosen to satisfy 
%$1/\tau_{LCR}\geqq5.8/\tau^{eff}_{TES}$ \cite{ref:tau_etf},
%where $tau_{LCR}$ is electrical time constant of $L/R_{bolo}$ .
%The frequency spacing is determined by the inductance of the LC filters.
%The criterion of the spacing are described in the following sections.

\subsection{Parasitic indcutance}
\subsubsection{SQUID and wiring impedance}

A strong constraint on the readout with the high bandwidth is
cross talk induced by the parasitic inductances 
of the SQUID input coil and the wiring.
They are shown at $L_{squid}$, $L_1$ and $L_2$ in Fig. \ref{fig:diagram_pb1} (a).
The current at resonant frequency $\omega_{i} = 1/\sqrt{LC_i}$ flows not only
through the $i$th bolometer but also into the $i\pm1$th channels. 
The voltage drop across the parasitic inductance modulates $i\pm1$ channels
and causes cross talk.
The power fluctuation of channel $i$ bolometer, $dP^{\omega_i}_{i}$, 
induces cross talk on the neighboring bolometers \cite{ref:afmux},

\begin{equation}
\frac{dP^{\omega_i}_{i\pm1}}{dP^{\omega_i}_i}
\simeq
-\frac{I^{\omega_i}_{i\pm1}}{I^{\omega_i}_{i}}
\frac{\omega_{i}}{\Delta \omega}
\frac{L_s}{L},
\label{eq:Lp}
\end{equation}

where $I^{\omega_i}_{i}$ is the current flowing through channel $i$ bolometer,
$L_s$ is sum of the parasitic inductance, $L_s = L_1 + L_2 + L_{squid}$,
and $\Delta \omega$ is the frequency spacing.
Equation (\ref{eq:Lp}) implies that the parasitic inductance should be
inversely proportional to the bandwidth to maintain the cross talk
at the same level.

The parasitic inductance $L_s$ also affects the bolometer stability.
%It spoils the voltage bias across the bolometer.
In the limit of infinite channel spacing, it shifts the resonant
frequency. At the resonant frequency $\omega_{i} = 1/\sqrt{(L + L_s)C_i}$, 
there is no effective series impedance. 
With finite spacing, $L_s$ cannot be removed completely due to the current
flowing through the neighboring channels.
The effective series impedance is approximately 
the ratio of parasitic inductance to channel spacing.
%For the neighboring channels $i \pm 1$, the current at frequency $\omega_{i}$
%is
%
%\begin{equation}
%I^{\omega_i}_{i\pm1}
%= \frac{V^{\omega_i}_{bias}}
%{R_{bolo} + j\omega (L+L_s) + 1/(j\omega_iC_{i\pm1})},
%\end{equation}
%
%\begin{equation}
%\simeq
%\frac{V^{\omega_i}_{bias}}
%{R_{bolo} + j2 \Delta \omega_i L},
%\end{equation}
%
%where $V^{\omega_i}_{bias}$ is a voltage bias applying across $LCR_{bolo}$
%cirucuit. The voltage drop across the parasitic inductance $V_s$ is approximately
%
%\begin{equation}
%V_s = j\omega_i L_s (I^{\omega_i}_{i-1} + I^{\omega_i}_{i+1}).
%\end{equation}
%
%An effective parasitic impedance $Z_s^{eff}$ is written by,
%\begin{equation}
%Z_s^{eff} = V_s / I_{bolo}
%= \frac{2j\omega_i L_s R_{bolo}^2}
%{R_{bolo}^2 + 4 \Delta \omega^2 L^2}.
%\end{equation}
If it becomes comparable with the bolometer resistance ($\sim1\Omega$),
it deteriorates the bolometer stability.
%For 90 kHz spacing, the effective impedance is 0.3 \% of the original
%inductance. 
The situation becomes worse as the carrier frequency increases.
Since the bandwidth used for POLARBEAR-2 readout is three times larger,
the parasitic inductance should be suppressed 
by 1/3 to avoid degradation of the performance.

\subsubsection{Suppression of parasitic inductance}
\begin{figure*}[hbt]
\centering
\subfigure[]{
\includegraphics[width=0.6\linewidth,keepaspectratio]{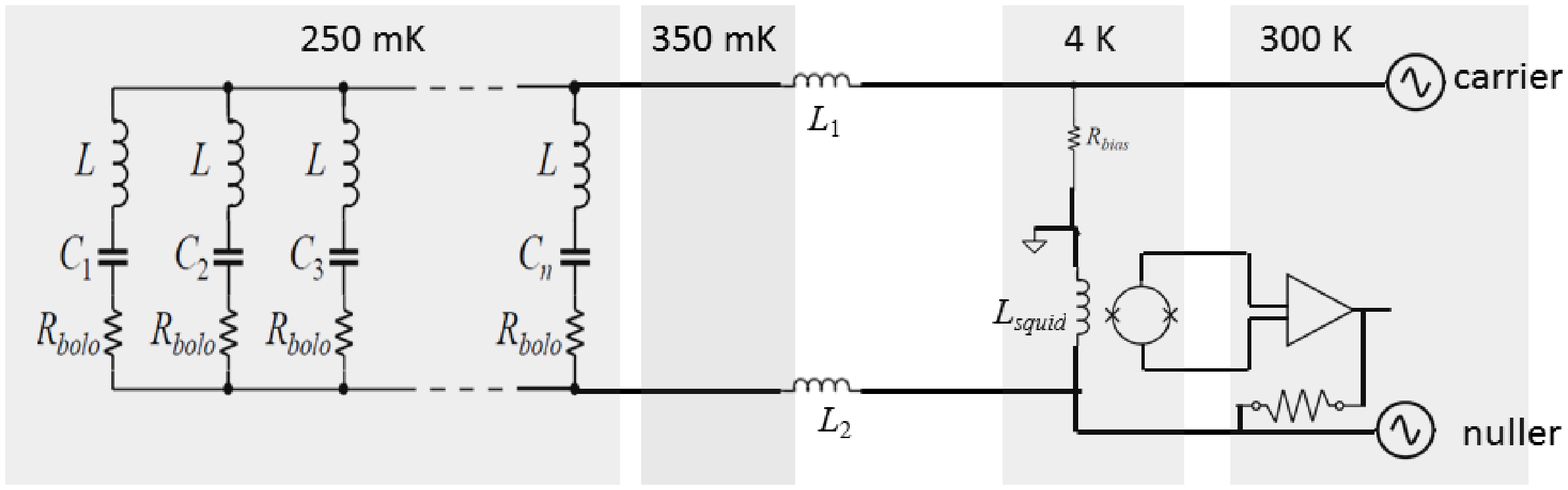}}
\subfigure[]{
\includegraphics[width=0.6\linewidth,keepaspectratio]{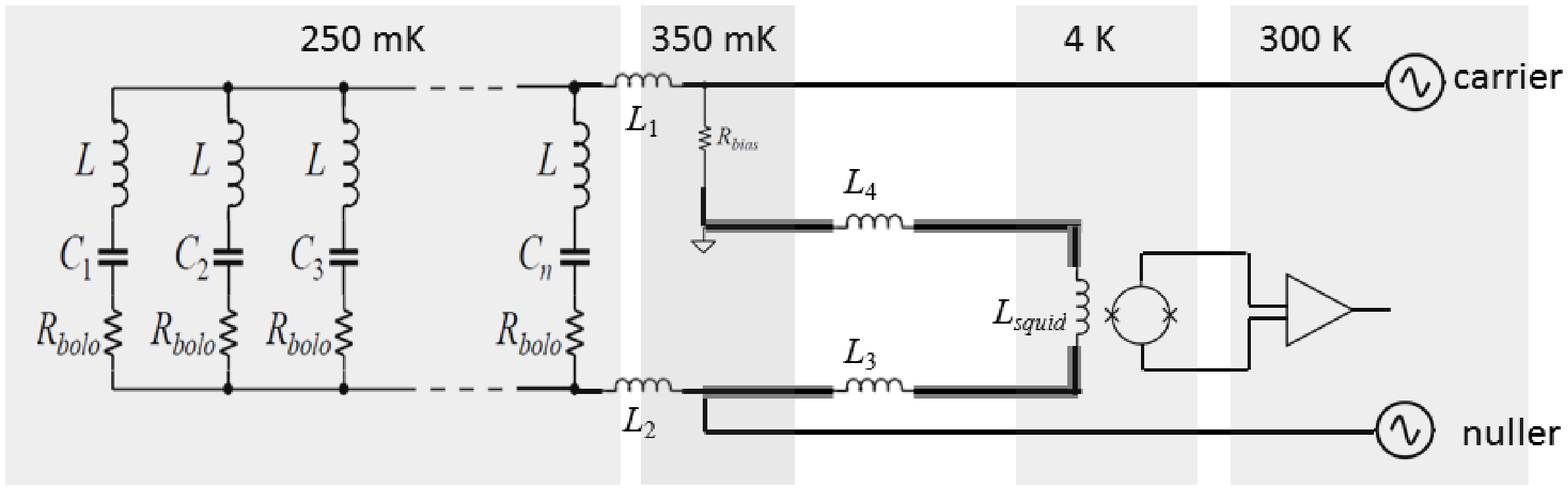}}
\caption{Block diagrams showing the digital fMUX readout of (a) POLARBEAR-1.
The resistor applying voltage bias across the bolometer is placed next the SQUID
at 4 K. The parasitic inductance of the wiring shown at $L_1$ and $L_2$
spoils the voltage bias. (b) For POLARBEAR-2, the resistor will move closer
to the $LCR_{bolo}$ circuit. The majority of the parasitic inductance of the
wiring shown at $L_3$ and $L_4$ is enclosed in the DAN loop, shown
at the highlighted line.}
\label{fig:diagram_pb1}
\end{figure*}

The parasitic inductance $L_1$ and $L_2$ shown in Fig. \ref{fig:diagram_pb1} (a)
arises from the cryogenic wiring that connects the bias resistor at 4 K
to the LC filters at 0.3 K.

For POLARBEAR-1, the bias resistor is mounted on the 4 K stage,
which is far from the 0.3 K stage where the LC filters are mounted.
The cable used for POLARBEAR-1 consists of two segments;
a 64-cm low inductance broadside coupled stripline 
which has high thermal conductivity, 
and a 10-cm high-inductance NbTi twisted pair which
is necessary to achieve thermal isolation between 4 K and 0.3 K.
Total parasitic inductance coming from the cable is 1.0 nH / cm
(stripline)$\times$64 cm $+$
7.0 nH / cm (twisted cables) $\times$10 cm $=$ 134 nH.
For POLARBEAR-2, it should be suppressed by 1/3, below 45 nH .
We have been developing a low inductance NbTi broadside coupled stripline
for thermal break. Assuming an inductance of 1.0 nH /cm
and the current cable length of 74 cm, 
the parasitic inductance would be 74 nH, which is much larger
%and length of 74 cm, the parasitic inductance will be 74 nH,
%which is much higher 
than the requirement.
The distance between the LC filters
and the bias resistor should be shorter than 45 cm.

To solve this issue, we will move the bias resistors 
closer to the $LCR_{bolo}$ circuit
and enclose the majority of the parasitic inductance in the DAN loop,
as shown in Fig. \ref{fig:diagram_pb1} (b).
DAN nulls current flowing in the DAN loop, shown in the highlighted
line in in Fig. \ref{fig:diagram_pb1} (b), effectively reducing
the parasitic inductance and the input impedance of the SQUID.
%The long SQUID input line is not applicable to  
%flux locked loop shunt feedback.
%The long wire produces phase shift along the feedback loop
%and makes the SQUID unstable.
The long SQUID input lines would not be suitable for the analog
Flux locked loop used in POLARBEAR-1. It can work well
with the DAN system, wherein the feedback occurs only 
in narrowband lines around the bolometer bias frequency.

Ideally, the bias resistor should be placed to next to the
$LCR_{bolo}$ circuit to minimize $L_1$ and $L_2$.
The position of the bias resistors is determined by the
cooling power of the cryostat.
Total power dissipation of the bias resistors will be 2.5 $\mu$W,
which is similar to the cooling power of the 0.25 K stage.
The bias resistor should be placed at 0.35 K, which have
more than 25 times cooling power in our cryogenic system. 
The bias resistor can be mounted on the NbTi broadside 
coupled stripline connecting 0.25 K and 4 K and
an be thermally linked to the 0.35 K stage by heat strap.
This setup can circumvent the limitation on the wiring length of 45 cm.
Another solution is to use a bias inductor that has no power
dissipation and can be mounted on the 0.25 K stage.
An advantage for this setup is that $L_1$ and $L_2$ will
be negligible. A disadvantage is that it will have three times
its number of cables running to the 0.25 K stage.
The thermal flow through these cables and the limited space
around the 0.25 K stage might be problematic.

%Another factor limiting the current bandwidth is parasitic inductance, 
%which we have reduced in our new readout architecture. 
%We will show frequency-domain multiplexing up to 3 MHz.

\subsection{Capacitors}
\subsubsection{Equivalent Series Resistance}

Another stray impedance that spoils the voltage bias across the bolometer
is the equivalent series resistance (ESR) 
of the capacitors that form the $LCR_{bolo}$ circuit.
Ideally, the ESR should be smaller than about 0.1 $\Omega$ ($0.1 R_{bolo}$). 
Assuming that the inductance $L$ is constant and the capacitance $C$
is used to select the frequency $f$,
the contribution of the dielectric loss to ESR is a linear function of frequency,
$R_{ESR} = 2\pi Lf \tan\delta$, where $\tan\delta$ is
the loss tangent of the dielectric material.
For the high frequency readout, material having low loss tangent should be used.
To have $R_{ESR} < 0.1 \Omega$ at 3 MHz, loss tangent should be $<2.2\times10^{-4}$.
For POLARBEAR-1, commercial surface mount multi-layer
ceramic capacitors were used. We investigated their usability
for POLARBEAR-2. 

\begin{figure}[hbt]
\centering
\includegraphics[width=0.6\columnwidth,keepaspectratio]{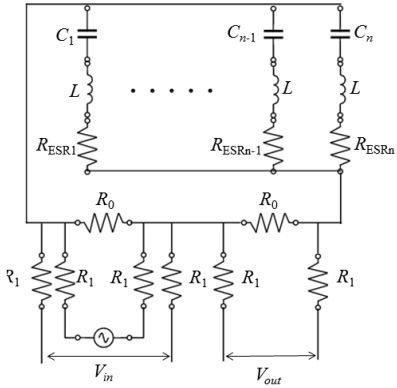}
\caption{Block diagram showing equivalent series resistance (ESR) measurement
setup.}
\label{fig:diagram_ESR}
\end{figure}

Figure \ref{fig:diagram_ESR} shows a block diagram
of ESR measurement setup. Each capacitor is connected to a
niobium inductor in series which forms LC filter.
%Since ESR depends on an inductance of the $LCR_{bolo}$ circuit,
We used 24-$\mu$H inductors that will be used for POLARBEAR-2.
The setup was cooled to 4.2 K using $^4He$.

At resonant frequency, the resistors $R_0$ and $R_{ESR}$ 
shown in Fig. \ref{fig:diagram_ESR}  form
a voltage divider. ESR is measured by frequency-sweeping 
a probe signal, monitoring $V_{in}$
and $V_{out}$, and measuring the peak height of each LC resonance.
To eliminate resistance of cables running from room temperature to
4 K, $R_1$, and $V_{out}/V_{in}$ are also measured.
The series resistance without a capacitor was 0.020 $\Omega$.

\begin{figure}[hbt]
\centering
\includegraphics[width=0.7\columnwidth,keepaspectratio]{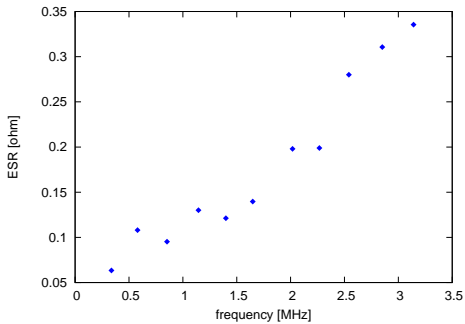}
\caption{Equivalent series resistance (ESR) of commercial surface mount multi-layer
ceramic capacitors measured at liquid $^4He$ temperature (4 K).}
\label{fig:ceramic_ESR}
\end{figure}

We have tested a wide variety of commercial surface mount multi-layer
ceramic capacitors. 
In Fig. \ref{fig:ceramic_ESR}, data points above 1 MHz show the best
ESR we have obtained so far. High-Q, low-loss capacitors
made by Vishay Vitramon and Johanson Techonology Inc. were
stacked in parallel to achieve the target capacitance.
The plots below 1 MHz are ESRs of the capacitors
used for POLARBEAR-1. 
As expected from the loss tangent, the ESR is high at high frequency.
The fluctuation of the ESRs are due to the stacked capacitors.
They were made from capacitors of different brands and processes, 
resulting different properties.
The ESR depends on not only the capacitance but also brands.
%Though these capacitors were from several
%brands and should have different loss tangent,
%ESR is approximately a linear function of frequency.
The ESR is acceptable below 1 MHz while it doesn't meet the requirement above 1 MHz.
We will investigate lower ESR capacitors.

\subsubsection{Capacitance accuracy}
It is important to set the bias frequencies with a constant spacing
in order to avoid cross talk, such as shown by Eq. (\ref{eq:Lp}).
Inductors we have used were fabricated at the National
Institute for Standards and Technology (NIST), with a
lithographic process. We are able to obtain sufficient uniformity
by lithography. With the high bandwidth readout, capacitance accuracy
is essential. At the highest frequency of 3 MHz, the accuracy
should be 0.6 \% assuming the error of the bias frequency is 10 \% (9 kHz).
The capacitance accuracy of commercial surface mount ceramic capacitors is
typically worse than 1 \%. Since most of channels have stacked capacitors in parallel,
the total error could be larger. 
For the best devices, the capacitance changes by several per cent or more after
being cooled from room temperature to sub-Kelvin.

To solve this issue, we will investigate lithographed capacitors.
The lithographic process is expected to control relative capacitances well.
Though the lithographed capacitors could have 
large deviation from the design capacitance value,
the absolute frequencies are much less important than the relative frequencies
in terms of cross talk. One candidate is parallel-plate capacitors
having superconducting plates separated by low loss tangent dielectric.
SRON has demonstrated such capacitors \cite{ref:SRON}.
Another candidate are interdigitated capacitors \cite{ref:interdigitated-c}
formed on a low loss material. Lumpled Element Kinetic Inductance Detectors
\cite{ref:LEKID}
having resonant frequencies of a few GHz utilize interdigitated capacitors,
which should have low loss.
Since there is no thin film involved in the design,
they will be less susceptible to ESD (electrostatic discharges),
which is often an issue with thin film capacitors.
%Lumpled Element Kinetic Inductance Detectors
%\cite{ref:LEKID}
%having resonant frequencies of a few GHz utilize interdigitated capacitors,
%which shoud have low loss.
%Since pattern of the interdigitated capacitors are formed
%on an insulator, they will have tolerance on ESD (electrostatic discharges),
%which is occasionaly an issue of thin film capacitors having small capacitances.

\section{High frequency Multiplexing testing}

\begin{figure}[hbt]
\centering
\includegraphics[width=0.7\columnwidth,keepaspectratio]{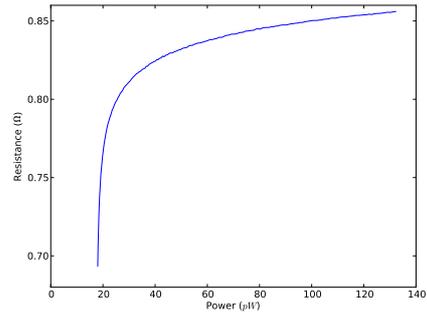}
\caption{Bolometer resistance vs. power for a bolometer voltage-biased at 3.10 MHz.}
\label{fig:IV_3.1}
\end{figure}

We have demonstrated high-frequency readout shown in Fig. \ref{fig:diagram_pb1} (b).
For LC filters, 24-$\mu$H inductors fabricated by NIST 
and the capacitors whose ESRs are shown in Fig. \ref{fig:ceramic_ESR} were used.
The majority of parasitic impedance is enclosed in the DAN loop.
The ratio is approximately $(L_1+L_2):(L_3+L_4)=1:25$.
%Resonant peaks of 13 $LCR_{bolo}$ channels are shown in Fig. \ref{fig:netanal}.
%The uneven spacing is due to change in capacitance at sub-Kelvin.
Figure \ref{fig:IV_3.1} shows bolometer resistance vs. power 
for a bolometer voltage-biased at 3.10 MHz.
The bolometer is operated dark at a bath temperature of 0.38 K.
As the bias voltage was reduced, the device fell into transition and
held the power applied across the bolometer by electro-thermal feedback (ETF).

%\begin{figure}[hbt]
%\centering
%\includegraphics[width=0.8\columnwidth,keepaspectratio]{netanal.eps}
%\caption{ The resonant frequencies of LC filters measured by frequency-sweeping 
%a small amplitude probe signal at 0.75 K where the bolometers are normal. }
%\label{fig:netanal}
%\end{figure}

\section{Future developments}

%We have shown that our system was able to bias bolometer in transition at 3 MHz.
We have shown that we can bias bolometers in their transition at 3 MHz.
To maintain the performance of TES bolometers at high frequency,
we will investigate low ESR capacitors. Capacitance accuracy will
%be important for the large focal plane to obtain high yield. 
be important for large focal planes 
such as that of POLARBEAR-2 to obtain high yield.
Lithographed capacitors will be expected to improve the yield.

We have been designing components for 32x multiplexing:
(1) a low inductance NbTi broad-side coupled stripline having a bias resistor on it,
and (2) a SQUID mounting board that has SQUIDs at 4 K. The new board will have
three pairs of wires between the 4K SQUID and
the sub-Kelvin bolometers; a carrier line, a nuller line and a return line.
This differs from the existing SQUID board that was used for POLARBEAR-1
which had only a single pair of wires providing the bolometer bias and return.
%three wires per SQUID, running to the cold stage, 
%carrier, nuller and SQUID input lines while the existing board
%used for POLARBEAR-1 has only a carrier line.
In future work, we will examine the performance of 
a high-bandwidth readout system with both the newly developed components 
and bolometers fabricated for the POLARBEAR-2 experiment.
%Integrating the newly developed components, we will examine the high-bandwidth readout
%with bolometers fabricated for POLARBEAR-2.

\section{Acknowledgement}
The POLARBEAR-2 experiment is funded by a Grant-in-Aid 
for Scientific Research from the Japan Society for the Promotion
of Science, and the National Science Foundation AST-1207892 of the United States of America.
The McGill authors acknowledge funding from the Natural Sciences 
and Engineering Research Council and Canadian Institute for Advanced Research.
All TES bolometers are fabricated at the UC Berkeley Microlab.
We thank NIST for design/fabrication of niobium inductors.

%\section{}
%\label{}



\end{document}